\documentclass[preprintnumbers,showpacs,amsmath,amssymb,floatfix,prd,onecolumn,superscriptaddress,nofootinbib]{revtex4}
\usepackage{graphicx}
\usepackage{epsfig}
\usepackage{bm}
\usepackage{amsfonts}
\usepackage{epstopdf}
\usepackage{bm}

\begin{document}

\title{  K-Inflation in Noncommutative Space-Time}

\author{Chao-Jun Feng}
\email{fengcj@shnu.edu.cn} 
\affiliation{Shanghai United Center for Astrophysics (SUCA), \\ Shanghai Normal University,
    100 Guilin Road, Shanghai 200234, P.R.China}
    \affiliation{State Key Laboratory of Theoretical Physics, \\Institute of Theoretical Physics, Chinese Academy of Sciences, Beijing 100190, P.R.China}

\author{Xin-Zhou Li}
\email{kychz@shnu.edu.cn} \affiliation{Shanghai United Center for Astrophysics (SUCA),  \\ Shanghai Normal University,
    100 Guilin Road, Shanghai 200234, P.R.China}

\author{Dao-Jun Liu}
\email{djliu@shnu.edu.cn} 
\affiliation{Shanghai United Center for Astrophysics (SUCA), \\ Shanghai Normal University,
    100 Guilin Road, Shanghai 200234, P.R.China}

\begin{abstract}
The power spectra of the scalar and tensor perturbations in the noncommutative k-inflation model are calculated in this paper. In this model, all the modes  created when the stringy space-time uncertainty relation is satisfied are  generated inside the sound/Hubble horizon during inflation for the scalar/tensor perturbations. It turns out that a linear term describing the noncommutative space-time effect contributes  to the power spectra of the scalar and tensor perturbations. Confronting the  general noncommutative k-inflation model  with latest results from \textit{Planck} and BICEP2, and taking $c_S$ and $\lambda$ as free parameters, we find that it is well consistent with observations. However, for the two specific models, i.e. the tachyon and DBI inflation models, it is found that  the DBI model is not favored, while the tachyon model lies inside the $1\sigma$ contour, if the e-folds number is assumed to be around $50\sim60$. 
\end{abstract}

 \pacs{98.80.Cq, 11.25.Wx}
\maketitle


\section{Introduction}
Inflation  \cite{Guth:1980zm,Linde:1981mu,Albrecht:1982wi} that solves a number of cosmological conundrums,  such as the horizon, monopole, entropy problems is now a crucial part of the history of our universe. Within a simplest inflation model, a field called inflaton plays a important role during the early time of the universe, which not only drives the universe to nearly exponentially expand, but also generates small fluctuations that is very important for large-scale structure formation in later time.  Inflation may be never end, which we call the eternal inflation \cite{Feng:2009kb, Feng:2010ya,Cai:2007et}.  From observations of the Cosmic Microwave Background (CMB), like the satellite-based Wilkinson Microwave Anisotropy Probe (WMAP) \cite{Hinshaw:2012aka} and  \textit{Planck} \cite{Ade:2013uln} experiments, we can obtain lots of informations about the early universe. One of them is the power spectra of the perturbations.

The observed CMB temperature fluctuations mainly generated by scalar perturbations already have constrain many inflation models, but there are still many left, which are all consistent with observations.  Large number of current CMB experiment efforts now target B-\textit{mode} polarization, which could be only generated by tensor perturbations. Recently, a ground-based  ``Background Imaging of Cosmic Extragalactic Polarization'' experiment has reported their results (BICEP2). They have shown that the observed B-\textit{mode} power spectrum at certain angular scales is well fitted by a lensed-$\Lambda$CDM + tensor theoretical model with tensor-to-scalar ratio $r=0.20^{+0.07}_{-0.05}$, and $r=0$ is disfavoured at $7.0\sigma$ \cite{Ade:2014xna}.

On the other side, general relativity might break down due to the very high energies during inflation, and as a candidate for the theory of everything, string theory should tell us what is a successful theory of the cosmology at that time, and  some corrections from string theory might be needed.  In the non-perturbative string/M theory, any physical process at the very short distance take an uncertainty relation, called  
the stringy space-time uncertainty relation (SSUR):
\begin{equation}\label{equ:ssur}
	\Delta t_p \Delta x_p \geq l_s^2 \,,
\end{equation}
where $l_s$ is the string length scale, and $\Delta t_p = \Delta t$, $\Delta x_p$ are the  uncertainties in the physical time and space coordinates.   The SSUR may be a universal relation for strings and D-branes \cite{yone, Li:1996rp,Yoneya:2000bt}.  Brandenberger and Ho \cite{Brandenberger:2002nq} have proposed a variation of space-time noncommutative field theory to realize the stringy space-time uncertainty relation without breaking any of the global symmetries of the homogeneous isotropic universe. If inflation is affected by physics at a scale close to string scale, one expects that space-time uncertainty must leave vestiges in the CMB power spectrum\cite{Huang:2003zp, Tsujikawa:2003gh,Huang:2003hw,Huang:2003fw,Liu:2004qe,Liu:2004xg,Cai:2007bw, Xue:2007bb}.  Recently, Feng \textit{et al.} \cite{Feng:2014yja} propose a  new power-law inflation model, by choosing a different $\beta_k^\pm$ functions (shall be defined below), which are equivalent to that proposed by  Brandenberger and Ho \cite{Brandenberger:2002nq}  by the means of integration. But it is much more clear to see the effect of noncommutative space-time and much easier to deal with the perturbation functions.  For detail calculations and discussions on it, see the Appendix~A in Ref.\cite{Feng:2014yja}. 

It has been shown that a large class of non-quadratic scalar kinetic terms can derive an inflationary evolution without the help of potential term, usually referred to as k-inflation \cite{ArmendarizPicon:1999rj, Garriga:1999vw}, e.g. the tachyon \cite{Sen:2002nu, Sen:2002in} and the Dirac-Born-Infeld (DBI) inflation \cite{Alishahiha:2004eh}. In this paper, we shall study the k-inflation model  in the noncommutative space-time following the method in Ref.\cite{Feng:2014yja}.  A  linear contribution to the power spectra of the scalar and tensor perturbations is given in this model. We also confront  two specific k-inflation models, namely the tachyon and DBI models, with latest results from the \textit{Planck} and BICEP2 experiments, and we  find that  the DBI model is not favored, while the tachyon model lies inside the $1\sigma$ contour, if the e-folds number is assumed to be around $50\sim60$.  This paper is organized  as follows. In next section, we will calculate the power spectra of the k-inflation model in the   noncommutative space-time; in Sec.\ref{sec:kinfmodels} we will constrain the parameters in tachyon and DBI models by using the latest observations. In the last section, we will draw our conclusions and give some discussions.

\section{K-inflation in noncommutative space-time} \label{sec:kind}

The general Lagrangian for a single field model with second-order field equations is an arbitrary function $p(\varphi, X)$ of the scalar field $\varphi$ and its kinetic energy  $ X = - \frac{1}{2}g^{\mu\nu}\nabla_\mu\varphi\nabla_\nu\varphi $. With the addition of gravity, the action takes the following form
\begin{equation}\label{equ:action}
	S = \int d^4x ~\sqrt{-g} ~\bigg[ ~\frac{R}{2} + p(\varphi, X) ~\bigg]\,,
\end{equation}
in the units of $M_{pl}^{-2} = 8\pi G =1$. The energy-momentum tensor of the inflaton reads
\begin{equation}\label{equ:emtensor}
	T_{\mu\nu} \equiv -\frac{2}{\sqrt{-g}} \frac{\delta S_\varphi}{\delta g^{\mu\nu}}  = p_{,X} \nabla_\mu\varphi\nabla_\nu\varphi + p g_{\mu\nu}\,, \quad p_{,X} \equiv \frac{\partial p(\varphi, X)}{\partial X} \,,
\end{equation}
which is equivalent to a perfect fluid, namely $T_{\mu\nu} = (\rho+p)U_\mu U_\nu + p g_{\mu\nu}$ with pressure $p$, energy density
\begin{equation}\label{equ:energydens}
	\rho = 2Xp_{,X} - p \,,
\end{equation}
and four-velocity $U_\mu = -\nabla_\mu \varphi / \sqrt{2X}$.  In the following, we will consider a flat  homogeneous and isotropic background model during inflation with the Friedmann-Robertson-Walker metric given by:
\begin{equation}\label{equ:FRW}
	ds^2 = -dt^2 + a^2(t) dx^2 \,.
\end{equation}
for a spatially flat universe ($K=0$).  Thus we have $X = \dot\varphi^2/2$  and the SSUR relation (\ref{equ:ssur}) becomes:
\begin{equation}\label{equ:ssur1}
	\Delta t \Delta x \geq \frac{l_s^2}{ a(t)} \,,
\end{equation}
which is not well defined when $\Delta t$ is large, because the argument $t$ for the scale factor on the r.h.s. changes over time interval $\Delta t$, and it is thus not clear what to use for $a(t)$ in Eq.(\ref{equ:ssur1}). The problem is the same when one uses the conformal time $\eta$ defined by $dt = ad\eta$. Therefore, for later use, a new time coordinate $\tau$ is introduced as
\begin{equation}\label{equ:tau}
	d\tau = a(t) dt \,,
\end{equation}
such that the metric becomes
\begin{equation}\label{equ:metric2}
	ds^2 = -a^{-2}(\tau) d\tau^2 + a^2(\tau)dx^2\,,
\end{equation}
and the SSUR relation is now well defined:
\begin{equation}\label{equ:ssur2}
	\Delta \tau \Delta x \geq l_s^2 \,.
\end{equation}
Therefore, the Friedmann equation becomes
\begin{equation}\label{equ:fried}
	H^2 = \frac{\rho}{3} = \frac{1}{3} ( \dot\varphi^2 p_{,X} - p )\,,
\end{equation}
and the equation of motion for the inflaton is
\begin{equation}\label{equ:eominf}
	\ddot\varphi~(p_{,X} + \dot\varphi^2 p_{,XX} ) + 3H \dot\varphi ~ p_{,X} = p_{,\varphi} - \dot\varphi^2~p_{,X\varphi} \,.
\end{equation}
Also we have the relation $\dot H = -X p_{,X} = -\dot\varphi^2p_{,X} /2$.  As a result of non-trivial kinetic terms in $p$, the dispersion relation for the inflaton is changed, and the fluctuations in the scalar field is no longer travelling at the speed of light. Instead, the sound speed in $\varphi$ is given by
\begin{equation}\label{equ:ss}
	c_s^2 \equiv \frac{p_{,X}}{\rho_{,X}} = \frac{p_{,X}}{p_{,X} + \dot\varphi^2 p_{,XX}}  \leq 1\,.
\end{equation}
In the canonical case, $p(\varphi, X)=X-V(\varphi)$ with inflaton potential $V(\varphi)$, we have $c_s^2 = 1$. To use the slow-roll approximation we may define 
\begin{equation}\label{equ:slowroll}
	\epsilon_1 = -\frac{\dot H}{H^2} = \frac{Xp_{,X}}{H^2}\,, \quad \epsilon_2 = \frac{\dot\epsilon_1}{\epsilon_1 H} \,,\quad s = \frac{\dot c_s}{c_s H} \,, 
\end{equation}
which are small parameters during inflation, i.e. $|\epsilon_1|, |\epsilon_2|, |s|\ll1$. 

\subsection{Scalar perturbation}

The action of the scalar perturbation could be written as
\begin{equation}\label{equ:pertaction}
	S =  \frac{V }{2}\int _{k<k_0}   d\eta d^3k ~ z^{2}(\eta) \bigg( \zeta'_{-k}\zeta'_{k} - c_s^2k^2 \zeta_{-k} \zeta_{k} \bigg)\,,
\end{equation}
where  $V$ is the total spatial coordinate volume and the prime denotes the derivatives with respect to a new time coordinate $\eta $  defined as
\begin{equation}\label{equ:teta1}
	\frac{d\eta}{d\tau} \equiv a^{-2}_{\text{eff}} =  \left( \frac{\beta_k^-}{\beta_k^+}\right)^{1/2}  = a^{-2}(\tau + \Delta \tau)\,.
\end{equation}
Here we have defined 
\begin{equation}\label{equ:beta1}
	\beta^{\pm}_k(\tau) =  a^{\pm2}\left(\tau+\Delta\tau\right) \,, \quad \Delta\tau = l_s^2k \,,
\end{equation}
where $\zeta_k$ is the curvature perturbation and $z \equiv a\sqrt{2\epsilon_1}/c_s$ is the so-called ``Mukhanov variable''.  Here we have taken a different form of the $\beta_k^\pm$ functions, which is equivalent to that used in the literatures by the mean of integration. It is now much more easier to deal with the equations of motion  for the field $\zeta_k$ :
\begin{equation}\label{equ:eom}
	u_k'' + \left( c_s^2k^2 - \frac{z''}{z} \right)u_k = 0 \,,
\end{equation}
which could be derived from the action (\ref{equ:pertaction}). Here the mode function is defined by $u_k = z \zeta_k$.   By using the definitions of slow-roll parameters, we get the coefficient of the third term in the perturbative equation (\ref{equ:eom})  as 
\begin{equation}\label{equ:z2z}
	\frac{z''}{z} \approx \frac{1}{\eta^2} (1-\epsilon_1)^{-2} \Sigma^{-2}(\tau,\Delta\tau)\bigg[ 2\Sigma(\tau,\Delta\tau) \left(1+\frac{\epsilon_2}{2} - s\right)  - \epsilon_1  +\frac{\epsilon_2}{2} - s \bigg] \,,
\end{equation}
where we have used 
\begin{equation}\label{equ:eta2h}
	\eta \approx - \bigg[ a(\tau+\Delta\tau)H(\tau+\Delta\tau) (1-\epsilon_1) \bigg]^{-1}\,,
\end{equation}
 derived from Eq.(\ref{equ:teta1}). Here, the $\Sigma$ function is defined as
\begin{equation}\label{equ:sig}
	\Sigma(\tau, \Delta\tau) = \frac{a(\tau)}{a(\tau+\Delta\tau)} \frac{H(\tau+\Delta\tau)}{H(\tau)} \,.
\end{equation}
By using the approximation $a(\tau + \Delta\tau)\approx a(\tau) + H(\tau)\Delta\tau$ and  $H(\tau + \Delta\tau)\approx H(\tau) - \epsilon_1 H^2(\tau) a^{-1}(\tau)\Delta\tau$, we get
\begin{equation}\label{equ:sig2approx}
	\Sigma(\tau, \Delta\tau) \approx 1-  (1+\epsilon_1)\frac{\lambda}{1-\lambda}+ \mathcal{O}(\Delta\tau^2)\,,
\end{equation}
where we have defined the parameter
\begin{equation}\label{equ:lambda}
	\lambda =  \frac{ \Delta\tau }{a(\tau)H^{-1}(\tau) + \Delta\tau}\,.
\end{equation}

All the modes  are created when the SSUR is saturated with the upper bound of the comoving wave number
\begin{equation}\label{equ:upbound}
	k_0(\tau) = \frac{a_{\text{eff}}}{l_s} = \frac{a(\tau + \Delta_\tau)}{l_s} \approx \frac{ H(\tau)}{l_s} \bigg[ a(\tau) H^{-1}(\tau) +\Delta\tau\bigg]\,,
\end{equation}
which means at time $\tau$, a mode with wave number $k_0$ is generated. The corresponding parameter $\lambda$ during inflation at the mode creating time is given by
\begin{equation}\label{equ:lambda0}
	\lambda_0 \equiv  \frac{ \Delta\tau }{a(\tau)H^{-1}(\tau) + \Delta\tau} \bigg|_{k=k_0}  \approx l_sH_* \,,
\end{equation}
where $H_*$ is the value of Hubble parameter during inflation.  Since  the scale factor is increasing  nearly exponentially ($a\sim e^{Ht}$) or  with a large power ($a\sim t^n$, with a large $n$) and $H_*$ is almost a constant, the parameter $\lambda$ will decrease with time, see Eq.(\ref{equ:lambda}).  At the same time when a mode is created (\ref{equ:upbound}),  the wave number cross the comoving sound horizon is given by
\begin{equation}\label{equ:cross}
	k_c = a(\tau) \frac{H(\tau)}{c_s} \,.
\end{equation}
Therefore, during inflation ,   we have
\begin{equation}\label{equ:k0kc}
	\frac{k_c}{k_0} \approx \frac{a(\tau)}{a(\tau) + H(\tau)\Delta\tau} \frac{l_s H(\tau)}{c_s} \approx   \frac{l_s H_*}{c_s}  \,.
\end{equation}
In the following, we will focus on the case of $l_sH_* /c_s \ll 1$, which means all the modes are created inside the horizon ($k_0\gg k_c$). The other case $l_sH_* /c_s \geq 1$ will not be considered in this paper, since in this case it is hard to explain the flatness of the universe, see Ref.\cite{Feng:2014yja} for detail discussions. Thus, during inflation the parameter $\lambda\ll1$ , and we will treat it as a free small parameter in the model. All the calculations are taken up to the first order of $\lambda$. Furthermore, the scalar power spectrum will be calculated at the time when the mode crosses the sound horizon ($k=aH/c_s$) in our case.   By using the approximation, Eq.(\ref{equ:eom}) becomes
\begin{equation}\label{equ:eom2}
	u_k'' + \left( c_s^2k^2 - \frac{\nu^2-1/4}{\eta^2} \right)u_k = 0 \,,
\end{equation}
where
\begin{equation}\label{equ:nu}
	\nu \approx   \frac{3}{2} + \epsilon_1 +\frac{\epsilon_2}{2}  - s  + \frac{2}{3}\lambda\,,
\end{equation}
up to the first order of slow-roll parameters and $\lambda$. With the initial Bunch-Davies vacuum condition:
\begin{equation}\label{equ:init}
	u_k = \frac{1}{\sqrt{2c_sk}} e^{-ic_sk\eta}\,,
\end{equation}
we get the solution to Eq.(\ref{equ:eom2}) 
\begin{equation}\label{equ:sol}
	u_k(\eta) = \frac{\sqrt{\pi}}{2} e^{i(\nu+1/2)\pi/2} \sqrt{-\eta} H_\nu^{(1)} (-c_sk\eta) \,,
\end{equation}
where $H_\nu^{(1)}$ is the Hankel's function of the first kind. At the superhorizon scales the solution becomes
\begin{equation}\label{equ:superhorizon}
	u_k(\eta) = 2^{\nu-3/2}  e^{i(\nu-1/2)\pi/2}  \frac{\Gamma(\nu)}{\Gamma(3/2)} \frac{1}{\sqrt{2c_sk}} (-c_sk\eta)^{1/2-\nu} \,.
\end{equation}
Therefore, the power spectrum of the metric scalar  perturbation is given by
\begin{equation}\label{equ:scalar}
	\mathcal{P}_s = \frac{k^3}{2\pi^2} |\zeta_k|^2  = \frac{k^3}{2\pi^2}\left |\frac{u_k}{z} \right|^2 
	=  \frac{2^{2\nu-4}  }{\epsilon_1c_s}  \left[ \frac{\Gamma(\nu)}{\Gamma(3/2)} \right]^2 \left(\frac{H}{2\pi} \right)^2\left(\frac{c_sk}{aH}\right)^{3-2\nu} \bigg|_{c_sk=aH} \approx \frac{1 }{8\pi^2\epsilon_1c_s}   \frac{H^2}{M_{pl}^2}  \bigg|_{c_sk=aH}\,,
\end{equation}
and the spectrum index of the power spectrum for the scalar perturbation reads
\begin{equation}\label{equ:sindex}
	n_s - 1 \equiv  \frac{d\ln \mathcal{P}_s}{d\ln k} = 3-2\nu - s \approx   - 2\epsilon_1 -\epsilon_2  + s  - \frac{4}{3}\lambda  \,.
\end{equation}
Also we  obtain the running of the index
\begin{equation}\label{equ:run}
	\alpha_s \equiv  \frac{ d n_s}{d\ln k}  = -2\epsilon_2\epsilon_1 -\epsilon_3\epsilon_2 +s_1s+\frac{4}{3}\lambda(1-\lambda)(1+\epsilon_1) \approx -2\epsilon_2\epsilon_1 -\epsilon_3\epsilon_2 +s_1s+\frac{4}{3}\lambda (1-\lambda+\epsilon_1)\,,
\end{equation}
where $\epsilon_3 \equiv \dot\epsilon_2/(\epsilon_2H)$ and $s_1 \equiv \dot s/(sH)$.

\subsection{Tensor perturbation}

The equation of motion for the tensor perturbation is almost the same as the one for the scalar perturbation, except that the Mukhanov variable becomes $z = a$. Then the equation of motion for the mode function is given by
\begin{equation}\label{equ:eomt}
	v_k'' + \left( k^2 - \frac{a''}{a} \right)v_k = 0 \,,
\end{equation}
where  $v_k\equiv a h_k/2$. Here $h_k$ denotes the independent degree of the tensor mode, i.e. $h_+$ and $h_\times$. By using the same approximation as that in the scalar perturbation, we get
\begin{equation}\label{equ:a2a}
	\frac{a''}{a} \approx \frac{1}{\eta^2} (1-\epsilon_1)^{-2} \Sigma^{-2}(\tau,\Delta\tau)\bigg[ 2\Sigma(\tau,\Delta\tau)  - \epsilon_1  \bigg] 
	\approx  \frac{\nu^2-1/4}{\eta^2}   \,,
\end{equation}
with
\begin{equation}\label{equ:nu2}
	\nu \approx  \frac{3}{2} + \epsilon_1+ \frac{2}{3}\lambda \,.
\end{equation}
Then the solution to Eq.(\ref{equ:eomt}) is given by
\begin{equation}\label{equ:sol}
	v_k(\eta) = \frac{\sqrt{\pi}}{2} e^{i(\nu+1/2)\pi/2} \sqrt{-\eta} H_\nu^{(1)} (-k\eta) \,,
\end{equation}
which also satisfies the Bunch-Davies vacuum initial condition. At the superhorizon scales the solution becomes
\begin{equation}\label{equ:superhorizont}
	u_k(\eta) = 2^{\nu-3/2}  e^{i(\nu-1/2)\pi/2}  \frac{\Gamma(\nu)}{\Gamma(3/2)} \frac{1}{\sqrt{2k}} (-k\eta)^{1/2-\nu} \,.
\end{equation}
Therefore, the power spectrum of the metric tensor perturbation is given by
\begin{equation}\label{equ:tensor}
	\mathcal{P}_t = 2\times\frac{k^3}{2\pi^2} |h_k|^2  = \frac{k^3}{\pi^2}\left |\frac{2v_k}{z} \right|^2 
	=  2^{2\nu}   \left[ \frac{\Gamma(\nu)}{\Gamma(3/2)} \right]^2 \left(\frac{H}{2\pi} \right)^2\left(\frac{k}{aH}\right)^{3-2\nu} \bigg|_{k=aH} \approx \frac{2}{\pi^2}   \frac{H^2}{M_{pl}^2}  \bigg|_{k=aH}\,,
\end{equation}
and the spectrum index of the power spectrum for the scalar perturbation reads
\begin{equation}\label{equ:tindex}
	n_t \equiv \frac{d\ln \mathcal{P}_t}{d\ln k} = 3-2\nu  \approx   - 2\epsilon_1  - \frac{4}{3}\lambda  \,.
\end{equation}
We also get the tensor-to-scalar ratio 
\begin{equation}
	r \equiv \frac{\mathcal{P}_t }{\mathcal{P}_s}=  16c_s\epsilon_1  \,,
\end{equation}
and the consistency relation 
\begin{equation}\label{equ:cr}
	r = - 8 c_s \left(n_t + \frac{4}{3}\lambda\right)  \,.
\end{equation}
When $\lambda\rightarrow0$, it reduces to the one in the commutative case, i.e. $r = -8c_sn_t$. With the help of $\lambda$ term in the above equation, one shall see that  the power-law inflation in noncommutative space-time may be more consistent with observations than that in the commnutative case.

\section{Confront two specific k-inflation models with observations} \label{sec:kinfmodels}

\subsection{Tachyon inflation models}

The tachyon inflation model was introduced by Sen \cite{Sen:2002nu, Sen:2002in}, and later studied in Ref.\cite{Gibbons:2002md,Kofman:2002rh,Fairbairn:2002yp,Piao:2002vf, Li:2002nk, Li:2002et,Li:2003ct,Liu:2004xg,Zhang:2010ym}.  In the following, we will consider the so-called assisted tachyon inflation mode \cite{Piao:2002vf}, which solves some difficulties with a single tachyonic field \cite{Kofman:2002rh}.  These difficulties could be also solved by taking account of non-minimal coupling to the gravity, see Ref.\cite{Piao:2002nh}. The Lagrangian for the assisted tachyon inflation reads
\begin{equation}\label{equ:tachyon}
	p(\varphi_i, X_i) = - \sum_{i=1}^{n}V(\varphi_{i})\sqrt{1-2l_s^2X_i} \,,
\end{equation}
which is a simply the sum of $n$ single-tachyonic fields without interactions. The $4$-d effective theory is applicable when $n$ is large enough, say $n\geq 10^7$ \cite{Piao:2002vf}. The Friedmann equation (\ref{equ:fried}) becomes
\begin{equation}\label{equ:trho}
	H^2 = \sum_{i=1}^{n}\frac{V(\varphi_i)}{3\sqrt{1-2l_s^2X_i}} = \sum_{i=1}^{n}\frac{V(\varphi_i)}{3\sqrt{1-l_s^2\dot\varphi_i^2}} \approx \frac{n}{3} V(\varphi_i) \,,
\end{equation}
and the equation of motion (\ref{equ:eominf}) for the inflaton is given by
\begin{equation}\label{equ:teominf}
	\ddot \varphi_i + 3H\dot\varphi_i (1-l_s^2\dot\varphi_i^2) + \frac{V_{,\varphi_i}}{Vl_s^2} (1-l_s^2\dot\varphi_i^2) = 0 \,.
\end{equation}
And also we have
\begin{equation}\label{equ:tdoth}
	\dot H = - \sum_{i=1}^{n}\frac{l_s^2\dot\varphi_i^2 V(\varphi_i)}{2\sqrt{1-l_s^2\dot\varphi_i^2}} \,, \quad \text{and} \quad  
	c_s^2 = 1-l_s^2\dot\varphi_i^2 \approx 1-\frac{2\epsilon_1 }{3n}\,.
\end{equation}
Thus, the slow-roll parameters (\ref{equ:slowroll}) are given by
\begin{eqnarray}\label{equ:tslowroll}
\epsilon_1 &=& \frac{3}{2}\sum_{i=1}^{n}l_s^2\dot\varphi_i^2 \approx \frac{1}{2l_s^2 nV(\varphi_i)} \left( \frac{V_{\varphi_i}}{V} \right)^2 
	\approx \frac{1}{2l_s^2 nV(\varphi_i)} \left( \frac{V_{,\varphi_i}}{V} \right)^2   \,,\\
\epsilon_2 &\approx& -\frac{2}{l_s^2nV(\varphi_i)} \frac{V_{,\varphi_i\varphi_i}}{V} + 6\epsilon_1 + \mathcal{O}(\epsilon^2)  \,,\\
s &=& \frac{\dot c_s}{c_s H} =  \frac{\dot c_s^2}{2c_s^2 H} \approx -\frac{\dot \epsilon_1}{3nc_s^2 H} \approx - \frac{\epsilon_2\epsilon_1}{3n}  \left( 1-\frac{1}{3n} \epsilon_1\right)^{-1} \sim \mathcal{O}(\epsilon^2) \,.
\end{eqnarray}
Considering the power-law potential $V(\varphi_i)=\alpha \varphi_i^m$, we have
\begin{equation}
	\epsilon_1 \approx \frac{m^2}{2l_s^2n V(\varphi_i)\varphi_i^2} \approx \frac{m}{2(m+2)N} \,, \quad 
	\epsilon_2 \approx  \frac{m(m+2)}{l_s^2 n V(\varphi_i)\varphi_i^2}  \approx \frac{1}{N} \,, \quad
	c_s \approx 1- \frac{\epsilon_1 }{3n} \,,
\end{equation}
where $N$ denotes the  number of e-folds during inflation, which  is defined by
\begin{equation}\label{equ:efold}
	N \equiv \int_t^{t_{\text{end}}} H dt \approx -\int_{\varphi_i}^{\varphi_i^{\text{end}}} 3H^2l_s^2\frac{V}{V_{,\varphi_i}} d\varphi_i 
	= - \frac{nl_s^2\alpha}{m} \int_{\varphi_i}^{\varphi_i^{\text{end}}} \varphi_i^{m+1} d\varphi_i 
	\approx  \frac{l_s^2n\alpha}{m(m+2)}  \varphi_i^{m+2}
	\approx  \frac{l_s^2 nV(\varphi_i) \varphi_i^{2}}{m(m+2)}   \,.
\end{equation}
Therefore, the spectrum indices of the power spectra for the scalar and tensor perturbations are given by
\begin{equation}\label{equ:tns}
	n_s -1= - \frac{2m+2}{(m+2)N} - \frac{4}{3}\lambda \,, \quad n_t = -\frac{m}{(m+2)N} - \frac{4}{3}\lambda \,,
\end{equation}
and the tensor-to-scalar ratio is
\begin{equation}\label{equ:tratio}
	r =\frac{8m}{(m+2)N} \left[1- \frac{m}{6n(m+2)N}\right] \approx  \frac{8m}{(m+2)N}\,.
\end{equation}
In fact, the sound speed $c_s$ is very closed to $1$ for  a positive value of $m$ when $n\geq10^7$ .

\subsection{DBI inflation models}
The Lagrangeian of DBI inflation is given by \cite{Alishahiha:2004eh} ( see also \cite{Silverstein:2003hf} ): 
\begin{equation}\label{equ:DBI}
	p(\varphi, X) = - \frac{1}{f(\varphi)}\sqrt{1-2f(\varphi)X} - V(\varphi) \,,
\end{equation}
where $f(\varphi)$ takes the form $f(\varphi) = \alpha/\varphi^4$. The Friedmann equation (\ref{equ:fried}) becomes
\begin{equation}\label{equ:drho}
	H^2 = \frac{1}{3} \left[ \frac{1}{f(\varphi) c_s } + V(\varphi)\right]\,,
\end{equation}
where the speed of sound from Eq.(\ref{equ:ss}) is given by
\begin{equation}\label{equ:dss}
	c_s = \sqrt{1-2f(\varphi)X} = \sqrt{1-f(\varphi)\dot\varphi^2} \,.
\end{equation}
While  the equation of motion (\ref{equ:eominf}) for the inflaton is 
\begin{equation}\label{equ:deominf}
	\frac{\ddot \varphi}{c_s^2} + 3H\dot\varphi + \frac{2\varphi^3}{\alpha}\bigg(3- \frac{1}{c_s^{2}}\bigg) + c_s V_{,\varphi}= 0 \,.
\end{equation}
And also we have
\begin{equation}\label{equ:tdoth}
	\dot H = -\frac{\dot\varphi^2}{2c_s} \,.
\end{equation}
Thus, the slow-roll parameters (\ref{equ:slowroll}) are given by
\begin{eqnarray}
\epsilon_1 &=& \frac{\dot\varphi^2}{2c_s H^2}  = \frac{1-c_s^2}{2c_s f(\varphi) H^2 }= \frac{3(1-c_s^2)}{2} \bigg[ 1+c_s f(\varphi)V(\varphi) \bigg]^{-1}
= \frac{3(1-c_s^2)}{2(1+\alpha' c_s)} \,,\label{equ:dslowroll1}\\
\epsilon_2 &=& -\frac{2c_s^2 s}{1-c_s^2} - \frac{s c_sfv}{1+c_sfV} - \sqrt{\frac{3c_s(1-c_s^2)}{1+c_sfV}} \frac{c_sfV}{1+c_sfV} (\ln fV)_{,\varphi}
= - \left( \frac{2c_s^2 }{1-c_s^2} + \frac{\alpha' c_s}{1+\alpha' c_s} \right)s  \,, \label{equ:dslowroll2}
\end{eqnarray}
where we have consider the quartic potential $V(\varphi) \sim \varphi^4$ that makes $f(\varphi)V(\varphi) = \alpha'$.  From Eq.(\ref{equ:dslowroll1}), we get $c_s \approx 1- (1+\alpha')\epsilon_1/3$ and the rate of change of the sound speed $s \approx -(1+\alpha')\epsilon_2\epsilon_1/3 \sim \mathcal{O}(\epsilon^2)$.  Then the Friedmann equation becomes
\begin{equation}
	H^2 = \frac{\varphi^4}{3\alpha c_s} (1+\alpha' c_s) \approx \frac{\varphi^4}{3\alpha} (1+\alpha')\left(1+\frac{\epsilon_1}{3}\right) \,.
\end{equation}
And the slow-roll parameters could be approximated as 
\begin{equation}\label{equ:dslowr}
	\epsilon_1 \approx \frac{8}{\varphi^2}  \approx \frac{1}{N}\,, \quad \epsilon_2 \approx \frac{8}{\varphi^2} \approx  \frac{1}{N}\,,
\end{equation}
where
\begin{equation}\label{equ:efold2}
	N = \int_{t}^{t_{\text{end}}} H dt = \int_{\varphi}^{\varphi^{\text{end}}}  \frac{H}{\dot\varphi}d\varphi \approx \frac{\varphi^2}{8} \,.
\end{equation}
Therefore, the spectrum indices of the power spectra for the scalar and tensor perturbations are given by
\begin{equation}\label{equ:dns}
	n_s -1= - \frac{3}{N} - \frac{4}{3}\lambda \,, \quad n_t = -\frac{2}{N} - \frac{4}{3}\lambda \,,
\end{equation}
and the tensor-to-scalar ratio is
\begin{equation}\label{equ:dratio}
	r = \frac{16}{N} \left[1-\frac{(1+\alpha')}{3N}\right] \approx \frac{16}{N}  \,.
\end{equation}
With the help of $\lambda$ term in the Eqs.(\ref{equ:tns}) and (\ref{equ:dns}), one shall see that  the tachyon and DBI inflation models in noncommutative space-time may be more consistent with observations than that in the commnutative case.

\subsection{Confront models with \textit{Planck} and BICEP2}

\begin{figure}
\centering
\includegraphics[width=0.4\linewidth]{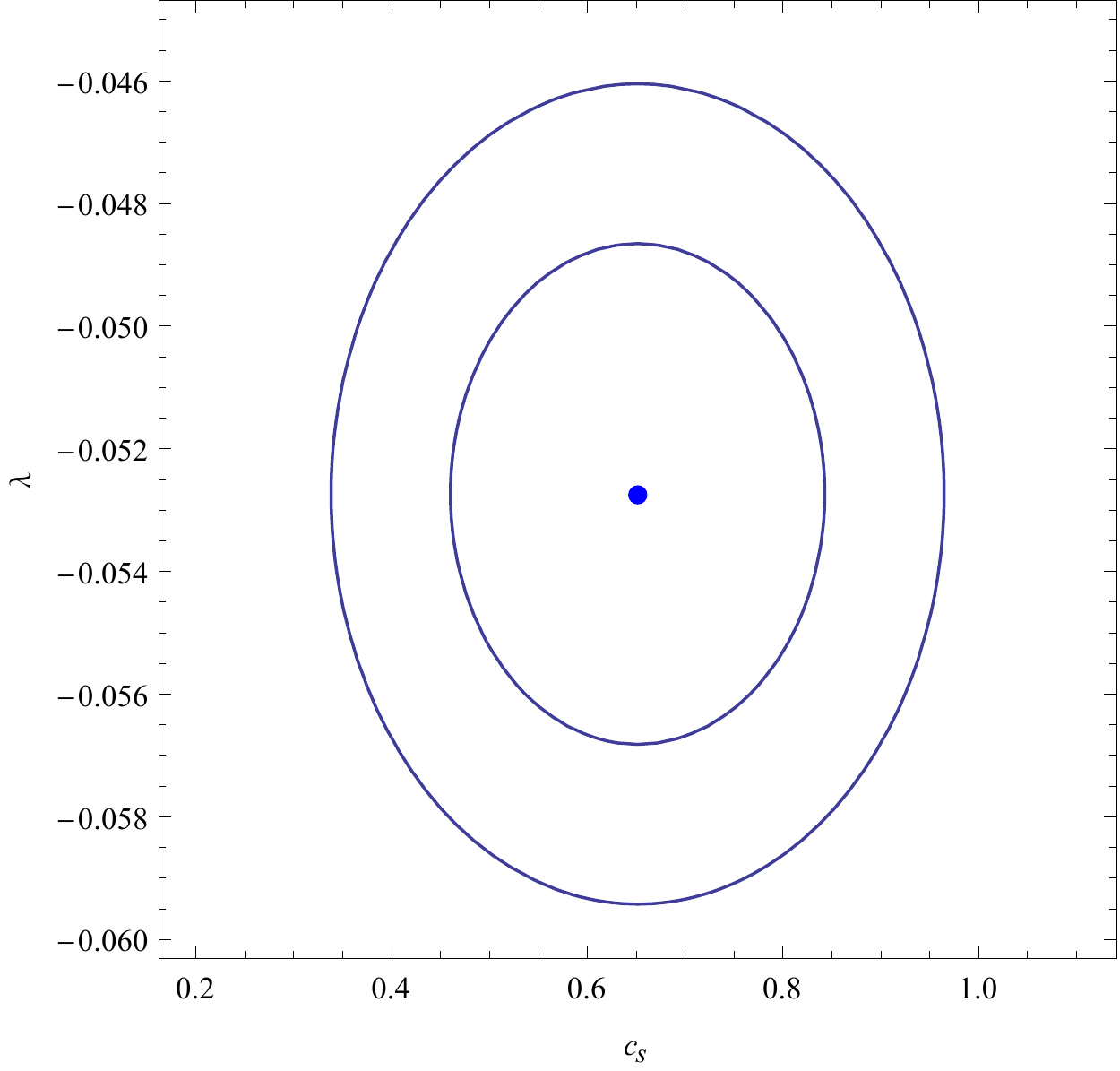}
\includegraphics[width=0.52\linewidth]{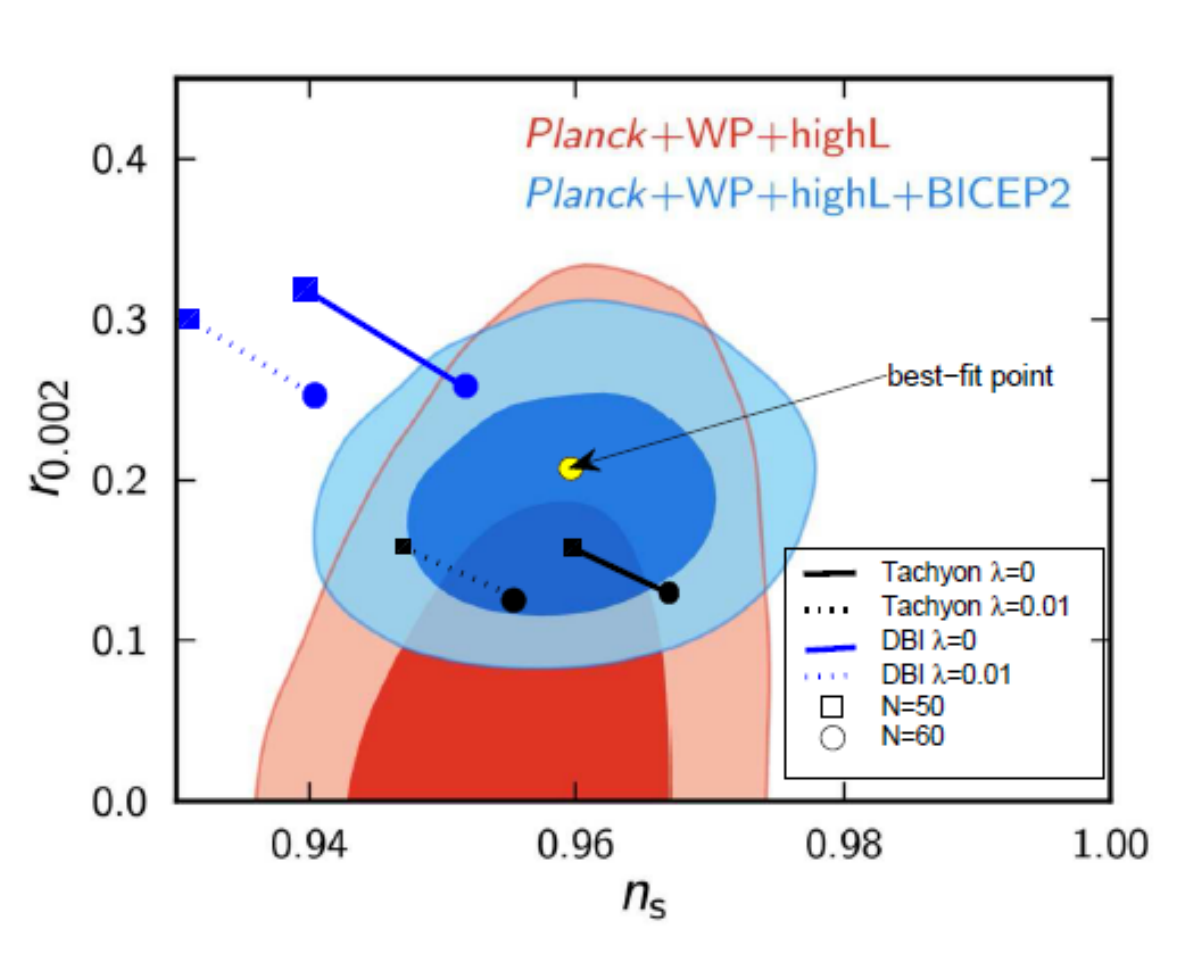}
\caption{Left: Constraints on the values of $c_s$ and $\lambda$. Two constraint contours are given at $68\%$ and $95\%$ confidence level. The central dot corresponds to  the best-fit point($c_s=0.65$, $\lambda=-0.0527$). Right: The $n_s-r$ plane based on Fig.13 from Ref.(\cite{Ade:2014xna}), in which the red contours are simply the MCMC result provided with the $Planck$ data release, while the blue ones are plotted when the BICEP2 data are added. The dark lines correspond to the tachyon inflation model with model parameter $m\rightarrow\infty$ and the blue lines correspond to DBI inflation model. The solid lines correspond to the model in usual commutative spacetime ($\lambda=0$), while the dotted ones correspond to the model in noncommutative spacetime with $\lambda=0.01$. The best-fit model for general k-inflation is also pointed out. }
\label{fig:fig1}
\end{figure}

In this subsection, we will constrain the noncommutative k-inflation by using the analyse results from data including the $Planck$ CMB temperature likelihood supplemented by the WMAP large scale polarization likelihood (henceforth $Planck$+WP). Other CMB data extending the \textit{Planck} data to higher-$l$, the \textit{Planck} lensing power spectrum, and BAO data are also combined, see Ref.\cite{Ade:2013uln} for details. In Ref.\cite{Ade:2013uln}, the index of scalar power spectrum is given by: $ 0.9583\pm0.0081$(\textit{Planck}+ WP), $ 0.9633\pm0.0072$(\textit{Planck}+WP+ lensing), $ 0.9570\pm0.0075$(\textit{Planck}+WP+highL), $ 0.9607\pm0.0063$(\textit{Planck} +WP+BAO). From the recent reports of BICEP2 experiment, we get the tensor-scalar-ratio as $r=0.20^{+0.07}_{-0.05}$, see Ref.\cite{Ade:2014xna} for details. Also, adopting the data from BICEP2 together with \textit{Planck}  and WMAP  polarization data, Cheng and Huang \cite{Cheng:2014ota} got the constraints of $r=0.23^{+0.05}_{-0.09}$, and  $n_t=0.03^{+0.13}_{-0.11}$. By using these results,  we obtain the constraints on the parameters $c_s$ and $\lambda$ as 
\begin{equation}\label{equ:constrain}
c_s=0.65\pm 0.19 \,,\quad \lambda=-0.0527\pm 0.041\,, \quad (68\% \text{CL}) \,.
\end{equation}
We plot the contours from $1\sigma$ to $2\sigma$ confidence levels for the parameters,  see Fig.\ref{fig:fig1}, in which the $n_s$-$r$ plane that based on Fig.13 from Ref.\cite{Ade:2014xna} is also presented.  From Fig.\ref{fig:fig1}, one can see that the general noncommutative k-inflation with its best fitting parameters is well consistent with observations, However, for the two specific models, if we assume that the number of e-folds number is around $50 \sim 60$, the DBI model is not favored, while the tachyon model lies inside the $1\sigma$ contour.

\section{Conclusions}
In conclusion, we have studied the k-inflation model  in the noncommutative space-time following the method in Ref.\cite{Feng:2014yja}.  A  linear contribution to the power spectra of the scalar and tensor perturbations is given in this model. We also confront  two specific k-inflation models, namely the tachyon and DBI models, with latest results from the \textit{Planck} and BICEP2 experiments, and we find that  the DBI model is not favored, while the tachyon model lies inside the $1\sigma$ contour, if the e-folds number is assumed to be around $50\sim60$.  We also constrained the parameter $c_s$ and $\gamma$ for a generic k-infaltion models, and find it is well-consistent with observations, see Fig.\ref{fig:fig1}.

\acknowledgments

This work is supported by National Science Foundation of China grant Nos.~11105091 and~11047138, ``Chen Guang" project supported by Shanghai Municipal Education Commission and Shanghai Education Development Foundation Grant No. 12CG51, National Education Foundation of China grant  No.~2009312711004, Shanghai Natural Science Foundation, China grant No.~10ZR1422000, Key Project of Chinese Ministry of Education grant, No.~211059,  and  Shanghai Special Education Foundation, No.~ssd10004, the Program of Shanghai Normal University (DXL124), and Shanghai Commission of Science and technology under Grant No.~12ZR1421700.


\begin{thebibliography}{999}

\bibitem{Guth:1980zm} 
  A.~H.~Guth,
  Phys.\ Rev.\ D {\bf 23}, 347 (1981).

\bibitem{Linde:1981mu} 
  A.~D.~Linde,
  Phys.\ Lett.\ B {\bf 108}, 389 (1982).
  
\bibitem{Albrecht:1982wi} 
  A.~Albrecht and P.~J.~Steinhardt,
  Phys.\ Rev.\ Lett.\  {\bf 48}, 1220 (1982).
  
\bibitem{Feng:2009kb} 
  C.~J.~Feng and X.~Z.~Li,
  Nucl.\ Phys.\ B {\bf 841}, 178 (2010)
  [arXiv:0911.3994 [astro-ph.CO]].
 
\bibitem{Feng:2010ya} 
  C.~J.~Feng, X.~Z.Li and E.~N.~Saridakis,
  Phys.\ Rev.\ D {\bf 82}, 023526 (2010)
  [arXiv:1004.1874 [astro-ph.CO]].

\bibitem{Cai:2007et} 
  Y.~F.~Cai and Y.~Wang,
  JCAP {\bf 0706}, 022 (2007)
  [arXiv:0706.0572 [hep-th]].


\bibitem{Hinshaw:2012aka} 
  G.~Hinshaw {\it et al.}  [WMAP Collaboration],
  Astrophys.\ J.\ Suppl.\  {\bf 208}, 19 (2013)
  [arXiv:1212.5226 [astro-ph.CO]].
  
\bibitem{Ade:2013uln} 
  P.~A.~R.~Ade {\it et al.}  [Planck Collaboration],
  arXiv:1303.5082 [astro-ph.CO].
  
\bibitem{Ade:2014xna} 
  P.~A.~R.~Ade {\it et al.}  [BICEP2 Collaboration],
  arXiv:1403.3985 [astro-ph.CO].
  
\bibitem{yone} 
T. Yoneya, in \textit{``Wandering in the Fields''}, eds. K. Kawarabayashi, A. Ukawa (World Scientific, 1987), P. 419;   
  
\bibitem{Li:1996rp} 
  M.~Li and T.~Yoneya,
  Phys.\ Rev.\ Lett.\  {\bf 78}, 1219 (1997)
  [hep-th/9611072].
  
\bibitem{Yoneya:2000bt} 
  T.~Yoneya,
  Prog.\ Theor.\ Phys.\  {\bf 103}, 1081 (2000)
  [hep-th/0004074].
  
\bibitem{Brandenberger:2002nq} 
  R.~Brandenberger and P.~M.~Ho,
  Phys.\ Rev.\ D {\bf 66}, 023517 (2002)
  [AAPPS Bull.\  {\bf 12N1}, 10 (2002)]
  [hep-th/0203119].
  
\bibitem{Huang:2003zp} 
  Q.~G.~Huang and M.~Li,
  JHEP {\bf 0306}, 014 (2003)
  [hep-th/0304203].
  
\bibitem{Tsujikawa:2003gh} 
  S.~Tsujikawa, R.~Maartens and R.~Brandenberger,
  Phys.\ Lett.\ B {\bf 574}, 141 (2003)
  [astro-ph/0308169].
  
\bibitem{Huang:2003hw} 
  Q.~G.~Huang and M.~Li,
  JCAP {\bf 0311}, 001 (2003)
  [astro-ph/0308458].
  
\bibitem{Huang:2003fw} 
  Q.~G.~Huang and M.~Li,
  Nucl.\ Phys.\ B {\bf 713}, 219 (2005)
  [astro-ph/0311378].
  
\bibitem{Liu:2004qe} 
  D.~J.~Liu and X.~Z.~Li,
  Phys.\ Lett.\ B {\bf 600}, 1 (2004)
  [hep-th/0409075].
  
\bibitem{Liu:2004xg} 
  D.~J.~Liu and X.~Z.~Li,
  Phys.\ Rev.\ D {\bf 70}, 123504 (2004)
  [astro-ph/0402063].
  
\bibitem{Cai:2007bw} 
  Y.~F.~Cai and Y.~Wang,
  JCAP {\bf 0801}, 001 (2008)
  [arXiv:0711.4423 [gr-qc]].
  
\bibitem{Xue:2007bb} 
  W.~Xue, B.~Chen and Y.~Wang,
  JCAP {\bf 0709}, 011 (2007)
  [arXiv:0706.1843 [hep-th]].

\bibitem{Feng:2014yja} 
  C.~J.~Feng, X.~Z.~Li and D.~J.~Liu,
  arXiv:1404.0168 [astro-ph.CO].

\bibitem{ArmendarizPicon:1999rj} 
  C.~Armendariz-Picon, T.~Damour and V.~F.~Mukhanov,
  Phys.\ Lett.\ B {\bf 458}, 209 (1999)
  [hep-th/9904075].
  
\bibitem{Garriga:1999vw} 
  J.~Garriga and V.~F.~Mukhanov,
  Phys.\ Lett.\ B {\bf 458}, 219 (1999)
  [hep-th/9904176].
  
\bibitem{Sen:2002nu} 
  A.~Sen,
  JHEP {\bf 0204}, 048 (2002)
  [hep-th/0203211].
  
\bibitem{Sen:2002in} 
  A.~Sen,
  JHEP {\bf 0207}, 065 (2002)
  [hep-th/0203265].
  
\bibitem{Alishahiha:2004eh} 
  M.~Alishahiha, E.~Silverstein and D.~Tong,
  Phys.\ Rev.\ D {\bf 70}, 123505 (2004)
  [hep-th/0404084].
  
\bibitem{Gibbons:2002md} 
  G WGibbons,
  Phys.\ Lett.\ B {\bf 537}, 1 (2002)
  [hep-th/0204008].
  
\bibitem{Kofman:2002rh} 
  L.~Kofman and A.~D.~Linde,
  JHEP {\bf 0207}, 004 (2002)
  [hep-th/0205121].
  
\bibitem{Fairbairn:2002yp} 
  M.~Fairbairn and M.~H.~G.~Tytgat,
  Phys.\ Lett.\ B {\bf 546}, 1 (2002)
  [hep-th/0204070].
  
\bibitem{Piao:2002vf} 
  Y.~S.~Piao, R.~G.~Cai, X.~M.~Zhang and Y.~Z.~Zhang,
  Phys.\ Rev.\ D {\bf 66}, 121301 (2002)
  [hep-ph/0207143].

\bibitem{Li:2002nk} 
  X.~Z.~Li, J.~G.~Hao and D.~J.~Liu,
  Chin.\ Phys.\ Lett.\  {\bf 19}, 1584 (2002)
  [hep-th/0204252].
  
\bibitem{Li:2002et} 
  X.~Z.~Li, D.~J.~Liu and J.~G.~Hao,
  hep-th/0207146.

\bibitem{Li:2003ct} 
  X.~Z.~Li and X.~H.~Zhai,
  Phys.\ Rev.\ D {\bf 67}, 067501 (2003)
  [hep-ph/0301063].
  
    
\bibitem{Zhang:2010ym} 
  H.~Zhang, X.~Z.~Li and H.~Noh,
  Phys.\ Lett.\ B {\bf 691}, 1 (2010)
  [arXiv:1006.2192 [hep-th]].

\bibitem{Piao:2002nh} 
  Y.~S.~Piao, Q.~G.~Huang, X.~M.~Zhang and Y.~Z.~Zhang,
  Phys.\ Lett.\ B {\bf 570}, 1 (2003)
  [hep-ph/0212219].

  
\bibitem{Silverstein:2003hf} 
  E.~Silverstein and D.~Tong,
  Phys.\ Rev.\ D {\bf 70}, 103505 (2004)
  [hep-th/0310221].


\bibitem{Cheng:2014ota} 
  C.~Cheng and Q.~G.~Huang,
  arXiv:1403.7173 [astro-ph.CO].
  
\bibitem{Ade:2013zuv} 
  P.~A.~R.~Ade {\it et al.}  [Planck Collaboration],
  arXiv:1303.5076 [astro-ph.CO].
  
    
\end{thebibliography}
\end{document}